\documentclass[sigconf]{acmart}

\usepackage{smartdiagram}
\usepackage{tikz}
\usepackage{pgfplots}
\usetikzlibrary{
    patterns,
}

\usepackage{booktabs}
\usepackage{bm}
\usepackage{caption}
\usepackage{subfigure}
\usepackage{color}
\usepackage{verbatim}
\usepackage{pdfpages}

\usepackage{algorithmicx}
\usepackage[ruled]{algorithm}
\usepackage{algpseudocode}

\usepackage[normalem]{ulem}

\definecolor{MyPink}{RGB}{255,178,178}
\definecolor{MyBlue}{RGB}{178,178,255}
\usetikzlibrary{er,positioning}

\setcopyright{acmcopyright}
\copyrightyear{2022}
\acmYear{2022}
\acmDOI{10.1145/1122445.1122456}

\copyrightyear{2023}
\acmYear{2023}
\setcopyright{acmlicensed}\acmConference[WSDM '23]{Proceedings of the
Sixteenth ACM International Conference on Web Search and Data
Mining}{February 27-March 3, 2023}{Singapore, Singapore}
\acmBooktitle{Proceedings of the Sixteenth ACM International Conference on Web Search and Data Mining (WSDM '23), February 27-March 3, 2023, Singapore, Singapore}
\acmPrice{15.00}
\acmDOI{10.1145/3539597.3570464}
\acmISBN{978-1-4503-9407-9/23/02}

\begin{document}

\title{Counterfactual Collaborative Reasoning}

\settopmatter{authorsperrow=4}

\author{Jianchao Ji}
\affiliation{
  \institution{Rutgers University}
  \city{New Brunswick, NJ}
  \country{US}
}
\email{jianchao.ji@rutgers.edu}

\author{Zelong Li}
\affiliation{ 
  \institution{Rutgers University}
  \city{New Brunswick, NJ}
  \country{US}
}
\email{zelong.li@rutgers.edu}

\author{Shuyuan Xu}
\affiliation{ 
  \institution{Rutgers University}
  \city{New Brunswick, NJ}
  \country{US}
}
\email{shuyuan.xu@rutgers.edu}

\author{Max Xiong}
\affiliation{ 
  \institution{Rutgers Preparatory School}
  \city{Somerset, NJ}
  \country{US}
}
\email{mxiong24@rutgersprep.org}

\author{Juntao Tan}
\affiliation{ 
  \institution{Rutgers University}
  \city{New Brunswick, NJ}
  \country{US}
}
\email{juntao.tan@rutgers.edu}

\author{Yingqiang Ge}
\affiliation{ 
  \institution{Rutgers University}
  \city{New Brunswick, NJ}
  \country{US}
}
\email{yingqiang.ge@rutgers.edu}

\author{Hao Wang}
\affiliation{ 
  \institution{Rutgers University}
  \city{New Brunswick, NJ}
  \country{US}
}
\email{hw488@cs.rutgers.edu}

\author{Yongfeng Zhang}
\affiliation{
  \institution{Rutgers University}
  \city{New Brunswick, NJ}
  \country{US}
}
\email{yongfeng.zhang@rutgers.edu}

\renewcommand{\shortauthors}{Jianchao Ji et al.}

\begin{abstract}
Causal reasoning and logical reasoning are two important types of reasoning abilities for human intelligence. However, their relationship has not been extensively explored under machine intelligence context. In this paper, we explore how the two reasoning abilities can be jointly modeled to enhance both accuracy and explainability of machine learning models. 
More specifically, by integrating two important types of reasoning ability---counterfactual reasoning and (neural) logical reasoning---we propose Counterfactual Collaborative Reasoning (CCR), which conducts counterfactual logic reasoning to improve the performance. In particular, we use recommender system as an example to show how CCR alleviate data scarcity, improve accuracy and enhance transparency. Technically, we leverage counterfactual reasoning to generate ``difficult'' counterfactual training examples for data augmentation, which---together with the original training examples---can enhance the model performance. Since the augmented data is model irrelevant, they can be used to enhance any model, enabling the wide applicability of the technique. Besides, most of the existing data augmentation methods focus on ``implicit data augmentation'' over users' implicit feedback, while our framework conducts ``explicit data augmentation'' over users explicit feedback based on counterfactual logic reasoning. Experiments on three real-world datasets show that CCR achieves better performance than non-augmented models and implicitly augmented models, and also improves model transparency by generating counterfactual explanations.

\end{abstract}



\maketitle

\section{Introduction}
Causal reasoning and logical reasoning are two important types of reasoning abilities. In this paper, we explore how the two reasoning abilities can be jointly modeled to enhance both accuracy and explainability of machine learning models. We take recommender system, which is an important prediction task, as an example to demonstrate the benefits of jointly modeling causal and logical reasoning. One important problem in recommendation model training is data scarcity. This is because user interactions are very sparse compared to the vast amount of items in the system, and this is especially true for sequential recommendation models which leverage a few or a few tens of user history interactions to predict the next interaction out of thousands or millions of candidates. One state-of-the-art approach to alleviating the data scarcity problem is counterfactual data augmentation, which conducts counterfactual reasoning over the original training examples to produce a set of informative augmented training examples \cite{wang2021counterfactual,yang2021top,fu2020counterfactual,chen2022data,xiong2021counterfactual,wen2020time}.
The augmented training examples, together with the original training examples, can help to improve the recommendation performance by generating synthetic data to cover the unexplored input space while maintaining correctness of the data as much as possible \cite{wen2020time}.

\begin{figure*}[t]
    \centering
    \includegraphics[scale=0.43]{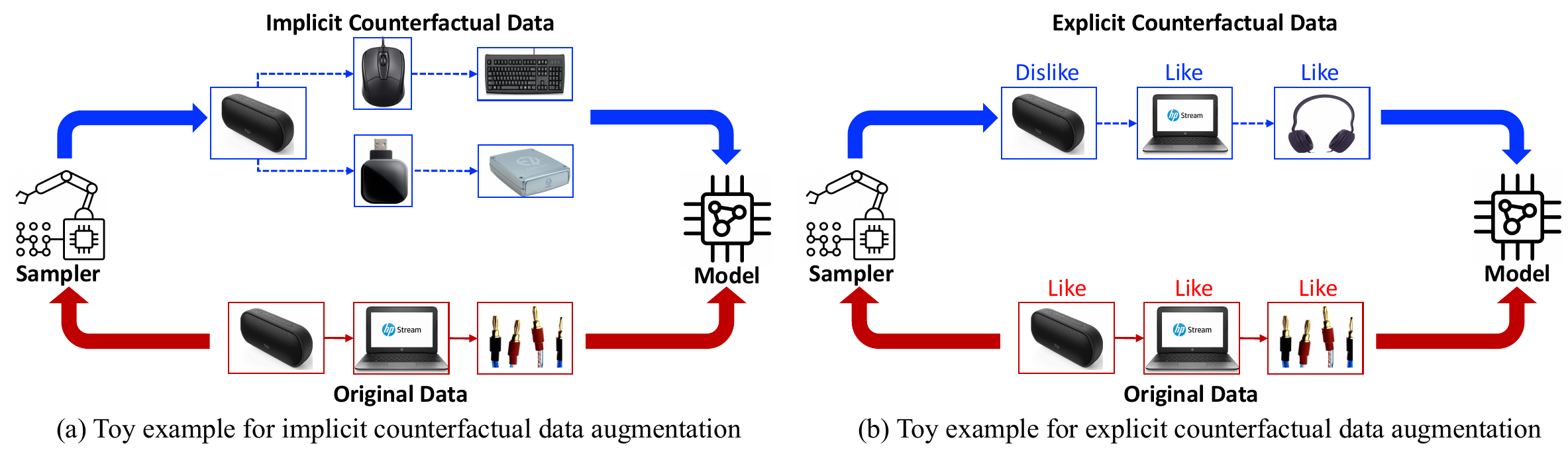}
    \vspace{-10pt}
    \caption{An illustration of (a) implicit counterfactual data augmentation and (b) explicit counterfactual data augmentation}
    \vspace{-10pt}
    \label{figure:toy_example}
\end{figure*}

Take Figure \ref{figure:toy_example}(a) as an example, the original training data shows that the user purchased a speaker, a laptop, and a data cable to connect the speaker to the laptop. Current data augmentation methods perturb one or more of the user's purchase histories to alternative items that can trigger the recommendation result to change. In this example, the laptop is perturbed to a mouse or a USB flash drive, and the recommendation result is changed to a keyboard or an SSD drive, respectively. Thus, two counterfactual examples are created. The key is to perturb the history item to very similar items that can change the output, which creates difficult examples to challenge the recommendation model and help train the model to distinguish such difficult examples for better performance \cite{wang2021counterfactual}.


Although counterfactual data augmentation has achieved success in improving many recommendation models, one major problem with existing counterfactual data augmentation methods is that they only consider users' implicit feedback such as click or purchase information for generating augmented counterfactual examples, while the explicit feedback such as users' like or dislike preference on the items are ignored. However, explicit feedback contains rich user preference information that helps to understand user behaviors and make accurate predictions.
Take Figure \ref{figure:toy_example}(b) as an example, which still shows the speaker, laptop and cable as the three items. If the user liked the speaker and laptop, then the user could indeed purchase the cable as the next item to connect the two. However, if the user disliked the speaker, then the next item could not be the cable but instead a headset as a substitute to the speaker. The example shows that even if the user's interaction history is the same, user's counterfactual preferences on the history items can create informative counterfactual examples to enhance the recommendation model. 

There is a reason that existing counterfactual data augmentation methods mostly focus on implicit data augmentation while ignoring the explicit feedback. Usually, counterfactual data augmentation relies on an anchor sequential recommendation model to perturb the history items and generate counterfactual examples. However, most of the existing sequential recommendation models work with implicit feedback and cannot handle explicit feedback in sequential modeling \cite{kang2018self,chen2018sequential,tang2018personalized,rendle2010factorizing,li2017neural,liu2018stamp,wang2021counterfactual,xiong2021counterfactual,hidasi2015session,he2016fusing,sun2019bert4rec,huang2018improving}. 
As a result, the counterfactual reasoning procedure is only able to perturb implicit feedback when generating counterfactual examples. Fortunately, recent advances on neural-symbolic/neural-logic reasoning methods \cite{shi2020neural,chen2021neural,hua2022system,chen2022graph,chen2022learn} such as neural collaborative reasoning (NCR) \cite{chen2021neural} shed light on this problem. By learning the logical inverse (NOT) operation, neural logical reasoning is able to model explicit feedback in sequential learning, thus making it possible to conduct counterfactual reasoning over users' explicit like/dislike feedback to generate explicit counterfactual examples for data augmentation.





In this paper, we integrate the two important types of reasoning ability---counterfactual reasoning and logical reasoning, and we propose a novel Counterfactual Collaborative Reasoning (CCR) framework which is able to generate explicit counterfactual examples to enhance the performance of sequential recommendation models. Technically, to create explicit counterfactual data, CCR provides a machine learning-based method to generate minimal changes on the users' historical explicit feedback that can lead to changes in the output under the NCR sampler model (Figure \ref{figure:toy_example}(b)). A very desirable feature of the CCR framework is its ``augment once, apply to all'' property, i.e., for a given dataset, we only need to conduct the data augmentation procedure for once to enrich the dataset, and the enriched dataset can be used to enhance the performance of multiple recommendation models. An additional benefit of the CCR framework is that the data augmentation procedure can naturally produce counterfactual explanations for the recommendation, which not only improves the recommendation performance but also helps to understand the reason of the recommendations. 

Experiments on three real-world datasets show that our CCR framework achieves significantly better performance than the models without data augmentation and the models with current existing data augmentation methods for sequential recommendation. Besides, quantitative evaluation results also show that our framework generates reliable explanations for the recommendations.

The key contributions of the paper are as follows:
\begin{itemize}
\item To the best of our knowledge, this work is the first to consider explicit counterfactual data augmentation for sequential recommendation. Besides, we demonstrate how logical reasoning and counterfactual reasoning--two of the most important reasoning abilities of humans--can be jointly modeled for better performance and explainability.
\item Under the ``augment once, apply to all'' framework, the generated explicit counterfactual data can improve the performance of multiple sequential recommendation models.
\item The data augmentation process not only enhances the recommendation performance but also improves the explainability.
\item We conduct experiments on several real-world datasets to analyze both the recommendation performance and the explanation performance.
\end{itemize}

The following part of the paper is organized as follows: We review related work in Section \ref{sec:related}, provide some preliminary knowledge in Section \ref{sec:preliminary}, present the details of our CCR framework in Section \ref{sec:framework}, and discuss the experimental results in Section \ref{sec:experiments}. We finally conclude the work with future research visions in Section \ref{sec:conclusions}.

\section{Related Work}
\label{sec:related}

Sequential recommendation is one of the most important types of recommendation models in real-world systems due to their generally good performance. The key idea is to predict the next item 
based on the user's historical interactions.
For example, Markov Chain-based models assume that each of the user's behavior is determined by the user's most recent behavior thus construct the transition matrix among items based on users' behaviors to predict users' future preference \cite{rendle2010factorizing,he2016fusing}.
To relax the most recent behavior assumption, researchers have made efforts to consider longer user behavior records for recommendation.
For example, recurrent neural network based models such as GRU4Rec \cite{hidasi2015session}, DREAM \cite{yu2016dynamic}, NARM \cite{li2017neural} and other variants \cite{donkers2017sequential,hidasi2018recurrent,quadrana2017personalizing,wu2017recurrent}
embed users' historical behaviors into a latent vector/representation to predict users' future behaviors. Recently, researchers also considered convolutional networks (Caser) \cite{tang2018personalized}, memory networks (MANN, KSR) \cite{chen2018sequential,huang2018improving}, attention mechanism (STAMP, SASRec) \cite{kang2018self,liu2018stamp}, self-supervised learning (S$^3$-Rec) \cite{zhou2020s3}, bi-directional transformers (BERT4Rec, SSE-PT, XLNet) \cite{sun2019bert4rec,wu2020sse,de2021transformers4rec}, logical reasoning (NCR) \cite{chen2021neural,shi2020neural,chen2022graph}, and foundation models (P5) \cite{geng2022recommendation} for sequential recommendation.

Compared with non-sequential models, sequential models need higher-quality sequential data for training since the output will be directly influenced by the input sequence \cite{wang2021counterfactual,chen2022data,xiong2021counterfactual}. However, the sparsity of real-world data may hinder the performance of the sequential recommendation models.
To alleviate this problem, researchers have been exploring counterfactual thinking for data augmentation so as to enhance both the dataset and the model. The data augmentation process uses alternatives to exchange the past behaviors and generate counterfactual examples \cite{epstude2008functional}. Following this idea, counterfactual data augmentation has made several important achievements in recommender systems \cite{wang2021counterfactual,yang2021top,xiong2021counterfactual,chen2022data,zhang2021causerec}.

However, existing counterfactual data augmentation methods for recommender systems only consider users' implicit feedback and cannot deal with the explicit feedback for counterfactual example generation \cite{wang2021counterfactual,yang2021top}. Actually, explicit information is also very useful in recommendation, since explicit and implicit feedback exhibit user preference from different perspectives. As a result, considering both types of feedback can enhance sequential modeling performance \cite{chen2018matrix}.
For example, neural logic reasoning (NLR) \cite{shi2020neural}, neural collaborative reasoning (NCR) \cite{chen2021neural}, and graph collaborative reasoning (GCR) \cite{chen2022graph} consider users' like and dislike feedback for sequential modeling.


Explainability is another important perspective for recommender systems research because it improves users' trust and satisfaction and helps the system designers in model debugging \cite{zhang2014explicit,zhang2020explainable,ge2022survey,chen2022measuring}. Researchers have explored various types of explanation styles, such as pre-defined templates \cite{zhang2014explicit,wang2018explainable}, image visualizations \cite{chen2019personalized,geng2022improving}, knowledge graph paths \cite{ai2018learning,xian2019reinforcement,geng2022path,chen2021temporal}, feature comparisons \cite{yang2022comparative,xian2021ex3}, reasoning rules \cite{shi2020neural,zhang2022neuro,chen2021neural,hua2022system,zhu2021faithfully,xian2020cafe,chen2022learn} and natural language sentences \cite{li2021personalized,geng2022recommendation,geng2022improving,li2022personalized,li2020generate,chen2019generate,li2017neural,li2021extra}, and recently, counterfactual reasoning has emerged as an effective method to generate explanations \cite{tan2021counterfactual,ghazimatin2020prince,tran2021counterfactual,xu2021learning}. Explainable AI has also been driving research in computer vision \cite{goyal2019counterfactual}, natural language processing \cite{feder2021causalm}, AI for Science \cite{li2022from,tan2022learning} and algorithmic fairness \cite{ge2022explainable,li2022fairness,fu2020fairness}.

\section{Preliminaries}
\label{sec:preliminary}
In this section, we briefly introduce the notations we used in this work as well as some background knowledge.

\subsection{Sequential Recommendation}
Suppose we have a user set $U=\{u_1,u_2,\cdots, u_{|U|}\}$ and an item set $V=\{v_1,v_2, \cdots, v_{|V|}\}$. The user $u_i$ interacted with a sequence of historical items $H_i=\{v_1^i,v_2^i, \cdots, v_n^i\}$. The corresponding feedback of these interacted historical items are $B_i=\{b_1^i,b_2^i, \cdots, b_n^i\}$. We let $b_t^i = 1$ if the user likes the item and $b_t^i = 0$ if the user dislikes the item. A sequential recommendation model $f$ predicts the ranking score $r_{ij}$ for user $u_i$ on item $v_j$ based on $H_i$ and $B_i$:
\begin{equation}
\begin{split}
    r_{ij} = f(H_i, B_i, v_j)
\end{split}
    \label{equation:1}  
\end{equation}
By ranking the candidate items in descending order of the ranking score $r_{ij}$, the recommendation model predicts the user's preferences and produces the recommendation list.
Sequential recommendation models can predict users' near future behaviors based on their history behaviors. Sometimes, we not only have users' interacted items from the dataset, but also we can know whether they like each item or not. However, most of the existing sequential recommendation models only consider the interacted histories for recommendation, or in other words, they only use implicit feedback information to train the model and make prediction, which may lose part of the useful information and may hinder the ability to make accurate predictions. One approach to modeling explicit feedback for sequential recommendation is neural collaborative reasoning (NCR) \cite{chen2021neural}, which we briefly introduce in the following.

\begin{figure}[t]
    \centering
    \includegraphics[scale=0.35]{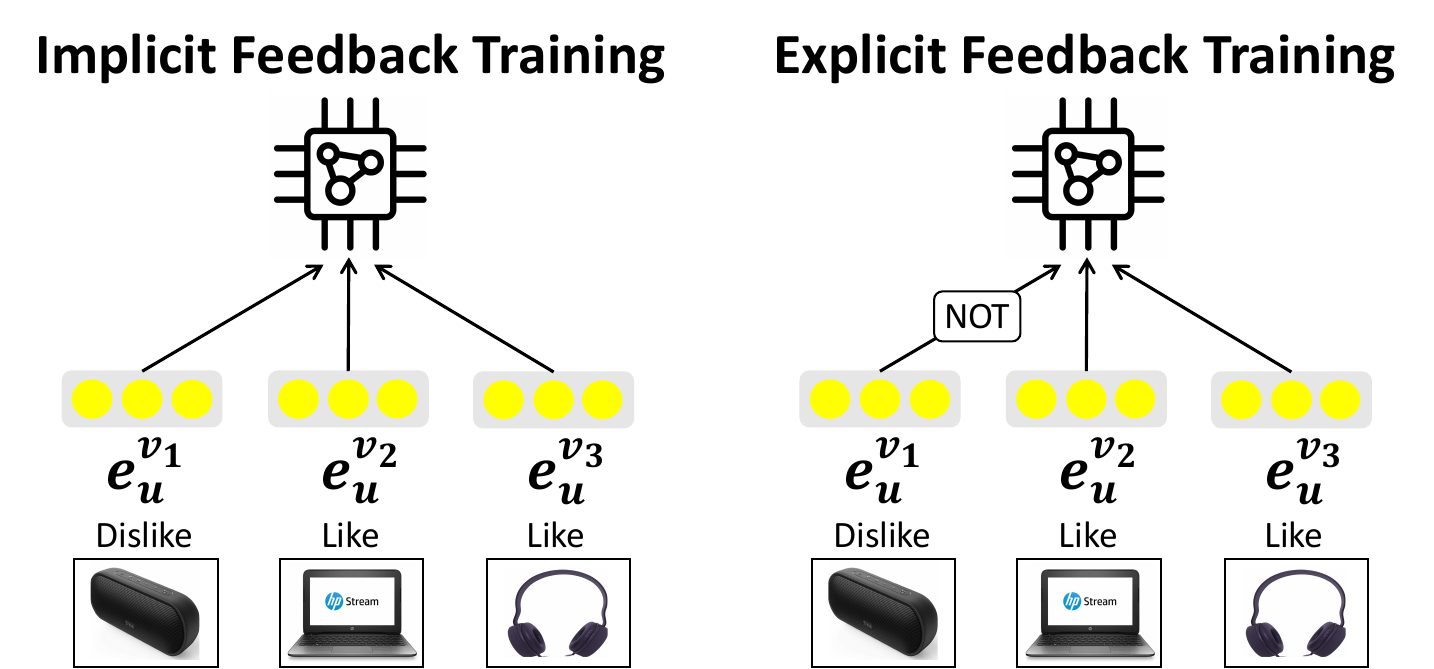}
    \vspace{-15pt}
    \caption{An illustration of the difference between sequential modeling with implicit feedback and with explicit feedback. The explicit feedback model leverages the logical NOT gate to distinguish between positive and negative feedback.}
    \label{figure: NCR} 
    \vspace{-15pt}
\end{figure}

\subsection{Sequential Modeling with Explicit Feedback}

Neural collaborative reasoning (NCR) \cite{chen2021neural} models the explicit feedback for sequential recommendation by training the neural logical NOT operator. By applying the NOT operation on history items, the model is able to distinguish users' positive and negative feedback on the historical items, as shown in Figure \ref{figure: NCR}. More specifically, NCR encodes a user-item interaction into an interaction event vector:
\begin{equation}
\begin{split}
    \bm{e_\bm{u}^\bm{v}} = \bm{W_2}\phi(\bm{W_1}\begin{bmatrix}\bm{u}\\\bm{v}\end{bmatrix} + \bm{b_1}) + \bm{b_2}
\end{split}
    \label{equation:2}  
\end{equation}
where $\bm{u},\bm{v}$ are user and item latent embedding vectors; $\bm{W_1},\bm{W_2}$ and $\bm{b_1},\bm{b_2}$ are weighted matrices and bias vectors that need to be learned; $\bm{e_\bm{u}^\bm{v}}$ is the encoded event vector that represents the interaction of user $u$ and item $v$, and $\phi(\cdot)$ is the activation function which is rectified linear unit (ReLU) in this work. By introducing neural logical modules: AND ($\wedge$), OR ($\vee$) and NOT ($\lnot$), NCR can transform the sequential data into a logical expression. Suppose user $u$'s feedback on item $v_1$ is negative and the feedback on other items are positive, then the explicit reasoning expression is:
\begin{equation}
\begin{split}
    \lnot\bm{e_\bm{u}^{\bm{v_1}}} \wedge \bm{e_\bm{u}^{\bm{v_2}}} \wedge \cdots \wedge \bm{e_\bm{u}^{\bm{v_n}}} \rightarrow \bm{e_\bm{u}^{\bm{v_{n+1}}}}
\end{split}
    \label{equation:3}  
\end{equation}
where $\{\bm{e_\bm{u}^{\bm{v_1}}}, \bm{e_\bm{u}^{\bm{v_2}}}, \cdots, \bm{e_\bm{u}^{\bm{v_n}}}\}$ represents the user's history event and $\bm{e_\bm{u}^{\bm{v_{n+1}}}}$ is the next item the user interacts with. To put explicit feedback information into consideration, $\bm{e_\bm{u}^\bm{v}}$ is used to represent that user $u$ interacted with item $v$ with positive feedback and and $\lnot\bm{e_\bm{u}^\bm{v}}$ is 
used to represent that user $\bm{u}$ interacted with item $v$ with negative feedback. Ideally, the sequential recommendation procedure can predict $\bm{e_\bm{u}^{\bm{v_{n+1}}}}$ based on $\{\bm{e_\bm{u}^{\bm{v_1}}}, \bm{e_\bm{u}^{\bm{v_2}}}, \cdots, \bm{e_\bm{u}^{\bm{v_n}}}\}$. Based on the definition of material implication\footnote{Material Implication ($\rightarrow$) can be represented as: $x \rightarrow y \Leftrightarrow \lnot x \vee y$}, the expression is equivalent to:
\begin{equation}
\begin{split}
    (\lnot\lnot\bm{e_\bm{u}^{\bm{v_1}}} \vee \lnot\bm{e_\bm{u}^{\bm{v_2}}} \vee \cdots \vee \lnot\bm{e_\bm{u}^{\bm{v_n}}}) \vee \bm{e_\bm{u}^{\bm{v_{n+1}}}}
\end{split}
    \label{equation:3}  
\end{equation}

The recommendation score of a candidate item $v_{n+1}$ is calculated based on the similarity between the logical expression and the constant True (T) vector. Based on the score, the model will decide whether the item should be recommend to the user (if the expression is close to True) or not (if the expression is close to False).

\section{Counterfactual Collaborative Reasoning (CCR)}
\label{sec:framework}

We build an counterfactual collaborative reasoning (CCR) framework to generate explicit counterfactual examples for data augmentation and improve the performance of sequential recommendation models. The main idea of the proposed data augmentation framework is to discover slight changes $\Delta$ on users' explicit feedback via solving a counterfactual optimization problem which will be formulated in the following. Meanwhile, the process of generating explicit counterfactual data can also provide explanations for items in the top-$K$ recommendation.

\subsection{Explicit Counterfactual Data Sampler}

As shown in figure \ref{figure:toy_example}(b), besides a sequential recommendation model $\bm{\mathcal{A}}$, our CCR framework introduces a sampler $\bm{\mathcal{S}}$ to generate explicit counterfactual examples. Firstly, both $\bm{\mathcal{A}}$ and $\bm{\mathcal{S}}$ in our model are pre-trained based on the original dataset. Then, the explicit counterfactual data generated by the sampler will be used to re-optimize the anchor model $\bm{\mathcal{A}}$. After that, the re-optimized anchor model will provide the final recommendation list for the user. 

To generate explicit counterfactual data, we use NCR as the sampler to conduct counterfactual reasoning.
This is because NCR can consider the counterfactual of explicit feedback with the help of logical negations ($\neg$). The first step of the sampler is to decide which explicit feedback of a user $u_i$'s historical items should be changed. We use a binary vector $\bm{\Delta_i}=\{0,1\}^{{\mid\bm{B_i}\mid}}$ to represent the intervention, where the the vector size is equal to the size of the user's explicit feedback vector $\bm{B_i}$. Then, we apply the intervention on $\bm{B_i}$:
\begin{equation}
\begin{split}
    \bm{B_i^*} = (1-\bm{B_i})\odot\bm{\Delta_i} + \bm{B_i}\odot(1-\bm{\Delta_i})
\end{split}
    \label{equation:5}  
\end{equation}

For each $\delta_t \in \bm{\Delta_i}$, if $\delta_t=1$, then the corresponding feedback is reversed; otherwise, the feedback remains the same. For example, if $\bm{B_i} = [0,1,1]$ and $\bm{\Delta_i} = [1,1,0]$, it means that the user's feedback on the first and second items should be reversed, thus $\bm{B_i^*} = [1,0,1]$. To decide which feedback should be changed, we design an optimization function for $\bm{\Delta_i}$:
\begin{equation}
\begin{split}
    \bm{\Delta_i} = {\text{argmin}}_{\bm{\Delta_i}} ~ \|\bm{\Delta_i}\|_0 + \, \alpha \cdot \bm{\mathcal{S}}(\bm{v_{n+1}}\mid \bm{H_i},\bm{B_i^*})
\end{split}
    \label{equation:6}  
\end{equation}
where $\|\bm{\Delta_i}\|_0$ is the zero-norm of the intervention vector $\bm{\Delta_i}$ that represents the amount of changed feedback, $\alpha$ is a hyper-parameter, and $\bm{v_{n+1}}$ is the item embedding vector of the ground-truth next item. $\bm{\mathcal{S}}(\bm{v_{n+1}}\mid\bm{H_i},\bm{B_i^*})$ is the ranking score of the sampler model for user $u_i$ on item $v_{n+1}$ under the counterfactual user feedback $\bm{B_i^*}$. In Eq.\eqref{equation:6}, the first term aims to minimize the amount of intervened feedback between the original data and the explicit counterfactual data. The second term tries to find the explicit feedback that can alter the output the sequence, i.e., the ranking score of item $v_{n+1}$ under the counterfactual feedback is decreased so that a new item appears as the output. However, the $\bm{\Delta_i}$ is not differentiable since it is discrete. Thus, we will introduce a relaxed optimization method later.

\begin{figure}[t]
    \centering
    \includegraphics[scale=0.6]{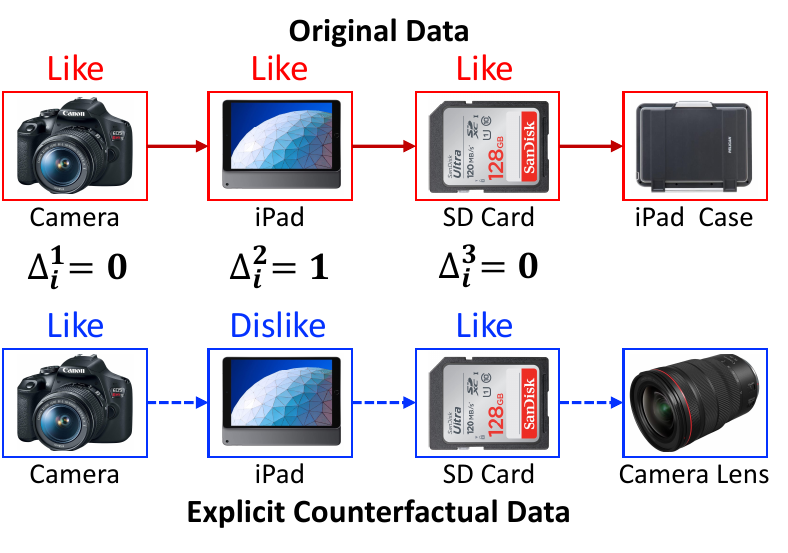}
    \vspace{-15pt}
    \caption{An illustration of explicit counterfactual data generation progress. In the original data, the user likes the iPad and then they purchased an iPad case which is compatible with the iPad. Suppose $\Delta_i^2 = 1$ and the user's explicit feedback on iPad is changed, then the sampler will generate a new item as the next item.}
    \label{figure: generation}
    \vspace{-15pt}
\end{figure}

To get the new next item, we take the history items $\bm{H_i}$ and the intervened feedback $\bm{B_i^*}$ into the sampler to derive the next item $\hat{\bm{v}}_{n+1}$ that the user may interact with:
\begin{equation}
\begin{split}
    \hat{\bm{v}}_{n+1} = \text{argmax}_{v \in I} ~ \bm{\mathcal{S}}(\bm{v}\mid \bm{H_i},\bm{B_i^*})
\end{split}
    \label{equation:7}  
\end{equation}
where $I$ is a set of items in the dataset, which can be the whole item set $V$ or another set involving prior knowledge. Finally, an explicit counterfactual sequence is generated as $(H_i,B_i^*,\hat{v}_{n+1})$. Together with the original sequence $(H_i,B_i,v_{n+1})$, the anchor model $\bm{\mathcal{A}}$ will be optimized over the augmented training dataset.

The intuition of the explicit counterfactual data augmentation procedure is that the sampler model creates ``difficult'' examples that leads to changes in the next item even if the historical feedback is just slightly changed. Such difficult examples, once included as augment examples to enhance the training data, will help the training of sequential recommendation models to distinguish the influence of minor changes in users' historical feedback \cite{wang2021counterfactual}. Take Figure \ref{figure: generation} as an example, in the original sequential data, user purchased an iPad case as the next item since it is compatible with the iPad that the user already purchased and the user actually likes the iPad. However, if the user dislike the iPad, then the next item would not be an iPad case but could be a camera lens since it is compatible with the camera that the user liked before.

\subsection{Relaxed Optimization}

One difficulty in optimizing Eq.\eqref{equation:6} is that the binary vector $\bm{\Delta_i}$ and the first term of the optimization function $\|\bm{\Delta_i}\|_0$ are not differentiable since they are discrete. In the implementation, we relax $\bm{\Delta_i}$ to a real-valued vector and relax the $\ell_0$-norm $\|\bm{\Delta_i}\|_0$ to $\ell_1$-norm $\|\bm{\Delta_i}\|_1$ to make the intervention vector $\bm{\Delta_i}$ learnable.
This method has been shown to be effective in prior research \cite{candes2005decoding,candes2006stable} and helps to minimize the number of changed explicit feedback in the sequence.
\begin{equation}
\begin{split}
    \bm{\Delta_i} = \text{argmin}_{\bm{\Delta_i}} \|\bm{\Delta_i}\|_1 + \, \alpha \cdot  \bm{\mathcal{S}}(\bm{v_{n+1}}\mid \bm{H_i},\bm{B_i^*})
\end{split}
    \label{equation:8} 
\end{equation}

As mentioned before, the sampler $\bm{\mathcal{S}}$ is implemented with NCR to accommodate the explicit feedback in the sequence. More specifically, for each user-item interaction event $e$ in the history $H_i$, suppose $\bm{e}$ is the corresponding event vector for the event (which could be negated if this event is a negative feedback in the original data), and suppose event $e$'s corresponding value in the intervention vector is $\delta_e$, then the intervened event vector $\bm{e}^*$ is:
\begin{equation}
    \bm{e}^* = \neg \bm{e} \cdot \delta_e + \bm{e} \cdot (1-\delta_e)
\end{equation}

These intervened event vectors constitute the counterfactual history $\{\bm{H_i},\bm{B_i^*}\}$ for the NCR sampler to calculate the ranking score of item $v_{n+1}$, as shown in Eq.\eqref{equation:8}.
After optimization, the values in the learned intervention vector $\bm{\Delta_i}$ may not be exactly equal to $0$ or $1$. As a result, a threshold will be applied to binarize $\bm{\Delta_i}$ as the final output. In the experiments, we set the threshold as $0.5$, i.e., for those elements in $\bm{\Delta_i}$ larger than $0.5$, we set them to $1$, otherwise, we set them to $0$. Finally, the binarized intervention vector $\bm{\Delta_i}$ is used to generate the new next item $\hat{v}_{n+1}$ for the explicit counterfactual example according to Eq.\eqref{equation:7}.


\subsection{Reduce Noisy Examples}
As mentioned above, for the generated explicit counterfactual data, the next interaction item is selected based on Eq.\eqref{equation:7}. However, since no sampler model is perfectly accurate, it may generate inaccurate predictions. Both of the accurate and inaccurate counterfactual examples generated by the sampler will be used to re-optimized the anchor model $\bm{\mathcal{A}}$, as a result, if we do not set some constraints to reduce the amount of inaccurate counterfactual examples, the performance of the re-optimized anchor model may be harmed. Inspired by \cite{wang2021counterfactual}, we set a confidence parameter $\kappa \in [0,1)$ to mitigate this issue. We accept the generated explicit counterfactual data only when $\bm{\mathcal{S}}(\hat{\bm{v}}_{n+1}\mid \bm{H_i},\bm{B_i^*}) > \kappa$. This means that we will only accept a counterfactual example when the sampler is sufficiently confident of it. Otherwise, the model will discard the example. 


\subsection{Learning Algorithm}

\setlength{\textfloatsep}{9pt}
\begin{algorithm}[t]
\caption{\mbox{CCR Learning Algorithm}}
\begin{algorithmic}
\State Input: the original dataset $T$
\State Input: Pre-train sampler $\bm{\mathcal{S}}$ and anchor model $\bm{\mathcal{A}}$
\State Initialize the counterfactual dataset $T_c = \emptyset$
\For{each training example from $T$}
\State Randomly initialize the intervention vector $\bm{\Delta_i}$
\State Learn $\bm{\Delta_i}$ by Eq.\eqref{equation:8}
\State Generate new sequence $(H_i,B_i^*,\hat{v}_{n+1})$ based on Eq.\eqref{equation:7}
\If{$\bm{\mathcal{S}}(\hat{\bm{v}}_{n+1}\mid \bm{H_i},\bm{B_i^*}) > \kappa$}
\State $ T_c \leftarrow T_c \cup (H_i,B_i^*,\hat{v}_{n+1})$
\EndIf
\EndFor
\State Re-optimize $\bm{\mathcal{A}}$ based on $T \cup T_c$
\State Output: the final recommendations based on re-optimized $\bm{\mathcal{A}}$
\end{algorithmic}
    \label{algorithm:ECDA} 
\end{algorithm}

To make the whole process clear, we summarize the learning algorithm of our framework in Algorithm \ref{algorithm:ECDA}. At first, we train both the sampler $\bm{\mathcal{S}}$ and the anchor model $\bm{\mathcal{A}}$ based on the original dataset $T$. Then for each training example in $T$,
the sampler will learn the intervention vector $\Delta_i$ and generate counterfactual data based on Eq.\eqref{equation:7}. If the sampler has enough confidence of the generated counterfactual data, it will be added into the counterfactual dataset $T_c$. When the model finishes the data augmentation process, the anchor model $\bm{\mathcal{A}}$ will be re-optimized based on $T_c\cup T$ to provide the final recommendations for each user.

\subsection{Counterfactual Explanations}
\label{sec:counterfactual_explanation}

During the process of generating explicit counterfactual data, our framework can also provide explanations to show why the model recommends the item to the user. Previous counterfactual explanation methods for recommendation \cite{tan2021counterfactual,ghazimatin2020prince,tran2021counterfactual} mostly focus on implicit counterfactual explanation based on implicit behaviors. However, one contribution of our work is that our framework can generate explicit counterfactual explanations.

\begin{figure}[t!]
    \centering
    \includegraphics[scale=0.6]{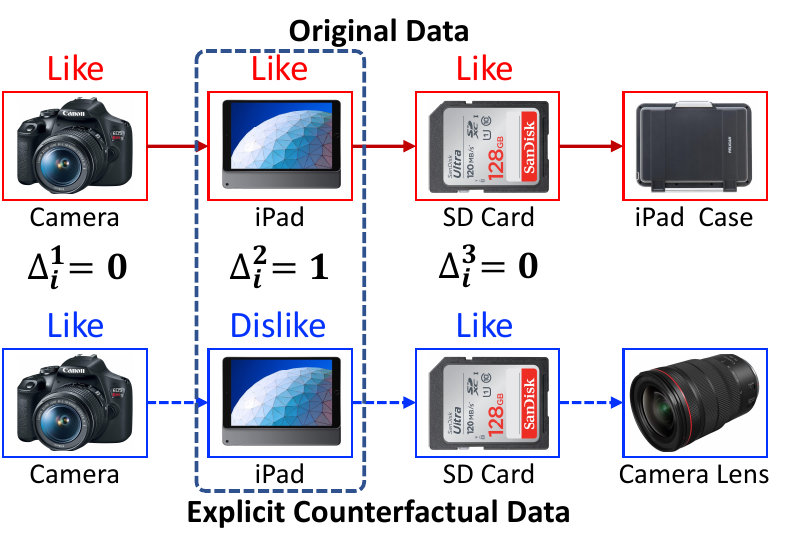}
    \vspace{-10pt}
    \caption{An example of generating counterfactual explanations. Since $\Delta_i^2 = 1$, the corresponding item is the counterfactual explanation for the recommended item.
    }
    \label{figure: explanation} 
    \vspace{-5pt}
\end{figure}

In Eq.\eqref{equation:8}, we are trying to explore an intervention vector $\bm{\Delta_i}$. Because of the first term in Eq.\eqref{equation:8}, only a few explicit feedback will be changed. Meanwhile, the second term of Eq.\eqref{equation:8} will penalize the probability of interacting with the current item. Therefore, only the most essential history items' feedback will be changed. These items can be used to generate counterfactual explanations for the recommended item. Take Figure \ref{figure: explanation} as an example, since $\Delta_i^2=1$, the corresponding item is the counterfactual explanation: the iPad is the reason for recommending the iPad case, because if the user disliked the iPad, we would not have recommended the iPad case but would have recommended the camera lens instead. We store all explanations in the set $E$.

To evaluate if our explanation correctly explains the recommended item, inspired by recent work on counterfactual explainable recommendation \cite{tan2022learning}, we use Probability of Necessity (PN) and Probability of Sufficiency (PS) to evaluate our explanations. In logic and mathematics, if $X$ happens then $Y$ will happen, we say $X$ is a sufficient condition for $Y$. Similarly, if $X$ does not happen then $Y$ will not happen, we say $X$ is a necessary condition for $Y$.

\subsubsection{\textbf{Probability of Necessity}}
Suppose a set of items $E_{ij} \subset V$ constitute the explanation for the recommended item $v_j$ to user $u_i$. The idea of the PN score is: 
if the items in $E_{ij}$ are reversed (for explicit explanation) or removed (for implicit explanation), then what is the probability that item $v_j$ would not be recommended for user $u_i$. We calculate the percentage of the generated explanations that meet the above PN condition:
\begin{equation}
\begin{split}
    \rm{PN} = \frac{\sum_{u_i\in U}\sum_{v_j\in R_{i,K}}\rm{PN}_{ij}}{\sum_{u_i\in U}\sum_{v_j\in R_{i,K}}I(E_{ij}\neq \emptyset)}
    , \; \rm{PN_{ij}}=\left \{ 
        \begin{array}{lr}
        1,~\rm{if} \; v_j \notin R_{i,K}^*\\
        0,~\rm{else}\\
        \end{array}
        \right.
\end{split}
    \label{equation:9}  
\end{equation}
where $\rm{R_{i,K}}$ is the original top-K recommendation list for user $u_i$. Let $v_j\in \rm{R_{i,K}}$ be a recommended item that our model has a nonempty explanation $E_{ij}\neq \emptyset$. Then for the original sequence data, we intervene (reverse or remove) the item(s) in $E_{ij}$ and get the new recommendation list $\rm{R_{i,K}^*}$ for user $u_i$ from the recommendation model. $I(E_{ij}\neq \emptyset)$ is an identity function: when $E_{ij}\neq \emptyset$, $I(E_{ij}\neq \emptyset)=1$. Otherwise, $I(E_{ij}\neq \emptyset)=0$.

\subsubsection{\textbf{Probability of Sufficiency}}

Similar to the definition of PN, the idea of PS score is: if the items in $E_{ij}$ are maintained while other items are reversed (for explicit explanation) or removed (for implicit explanation), 
then what is the probability that item $v_j$ would still be recommended for user $u_i$. We calculate the percentage of the generated explanations that meet the above PS condition:
\begin{equation}
\begin{split}
    \rm{PS} = \frac{\sum_{u_i\in U}\sum_{v_j\in R_{i,K}}\rm{PS_{ij}}}{\sum_{u_i\in U}\sum_{v_j\in R_{i,K}}(E_{ij}\neq \emptyset)}
    ,\; \rm{PS_{ij}}=\left \{ 
        \begin{array}{lr}
        1,~\rm{if} \; v_j \in R_{i,K}^{'}\\
        0,~\rm{else}\\
        \end{array}
        \right.
\end{split}
    \label{equation:10}  
\end{equation}
where $\rm{R_{i,K}^{'}}$ is the new recommendation list after the intervention is applied, and other notations have similar meanings as above.

\section{Experiments}
\label{sec:experiments}

In this section, we conduct experiments on three real-world datasets and compare the results of (1) the original sequential recommendation model without data augmentation, (2) models with implicit counterfactual data augmentation, and (3) models with our Counterfactual Collaborative Reasoning (CCR) framework. 
Furthermore, the counterfactual explanation results show our framework's ability to generate higher quality explanations.

\subsection{Dataset}
We use three real-world datasets in the experiments. 

\textbf{ML100K} \cite{harper2015movielens}: The MovieLens-100K (ML100K) dataset 
stores users' preference for various movies. It contains 100,000 movie ratings from 1 to 5 stars of 943 users to 1,682 movies.

\textbf{Amazon} \cite{ni2019justifying}: This is the Amazon e-commerce dataset.
We take \textbf{Movies $\&$ TV} and \textbf{Electronics} datasets as two examples for experiments. Movies $\&$ TV contains 123,961 users, 50,053 products and 1,697,533 product ratings. Electronics contains 192,404 users, 63,002 products and 1,689,188 product ratings. 

Some basic statistic of the datasets can be found in Table \ref{Table:dataset}.
We consider 1-3 ratings as negative feedback with label as 0, and 4-5 ratings as positive feedback with label as 1.
We use positive Leave-One-Out \cite{chen2021neural,zhao2020revisiting} to create the training, validation and testing dataset. For each user, we put the last positive interaction and its following negative interactions into the testing set, and we put the last but one positive interaction and its following negative interactions into the validation set. Then, we put all of the rest interactions into the training set. If a user has less than 5 interactions, we put all of the interactions into the training set to avoid cold-start.

\subsection{Baselines}
We consider both standalone sequential recommendation models and implicit data augmentation methods for comparison:

\textbf{GRU4Rec} \cite{hidasi2015session}: GRU4Rec is a sequential recommendation model based on Recurrent Neural Networks (RNN).

\textbf{STAMP} \cite{liu2018stamp}: STAMP is a sequential recommendation model based on attention mechanism, which can capture users' long-term and short-term preferences for recommendation.

\textbf{SASRec} \cite{kang2018self}: SASRec is a sequential recommendation model based on self-attention mechanism

\textbf{NCR} \cite{chen2021neural}: NCR is a sequential recommendation model based on neural logical reasoning, which captures the logical relationship between user-item interactions for recommendation.

\textbf{CASR} \cite{wang2021counterfactual}: CASR is a state-of-the-art implicit counterfactual data augmentation method for sequential modeling. 

Both CASR and our CCR frameworks can be applied on all of the four recommendation models.


\begin{table}[t!]
    \centering
    \begin{tabular}{l|rrrr}
        \toprule
         Dataset & \#users & \#items & \#interaction & Density \\
         \midrule
         ML100K & 943 &1,682 &100,000 &6.30\% \\
         Movies \& TV  &123,961 &50,053 &1,697,533 &0.027\%\\
         Electronics  &192,404  &63,002 &1,689,188 &0.014\%\\
         \bottomrule
    \end{tabular}
    \caption{Basic statistics of the datasets}
    \label{Table:dataset}
    \vspace{-25pt}
\end{table}

\begin{table*}[t]
    \centering
    \setlength{\tabcolsep}{5pt}
    \begin{tabular}{r|c|c|c|c|c|c|c|c|c|c|c|c}
    \toprule
         Dataset & \multicolumn{4}{|c|}{ML100K} & \multicolumn{4}{c|}{Movies \& TV}& \multicolumn{4}{c}{Electronics}\\
         \midrule
         Metric & \footnotesize{NDCG@5}& \footnotesize{NDCG@10} & \footnotesize{HR@5}  &\footnotesize{HR@10} &\footnotesize{NDCG@5} &\footnotesize{NDCG@10} & \footnotesize{HR@5} &\footnotesize{HR@10} &\footnotesize{NDCG@5} &\footnotesize{NDCG@10} & \footnotesize{HR@5} &\footnotesize{HR@10}  \\
         \midrule
         STAMP &0.342 &0.402 &0.503 &0.665 &0.406 &0.427 &0.521 &0.657 &0.301 &0.341 &0.412 &0.542\\
         CASR-STAMP &0.351 &0.406 &0.511 &0.676 &0.412 &0.445 &0.532 &0.661 &0.307 &0.349 &0.425 &0.553\\
         CCR-STAMP &$\bm{0.365^*}$ &$\bm{0.418^*}$ &$\bm{0.527^*}$ &$\bm{0.693^*}$ &$\bm{0.435^*}$ &$\bm{0.463^*}$ &$\bm{0.552^*}$ &$\bm{0.689^*}$ &$\bm{0.329^*}$ &$\bm{0.364^*}$ &$\bm{0.453^*}$ &$\bm{0.570^*}$\\
         \midrule
         GRU4Rec &0.340 &0.403 &0.502 &0.672 &0.411 &0.431 &0.538 &0.661 &0.312 &0.354 &0.432 &0.554\\
         CASR-GRU4Rec &0.349 &0.411 &0.509 &0.680 &0.414 &0.453 & 0.542 &0.671 &0.326 &0.369 &0.447 &0.560\\
         CCR-GRU4Rec &$\bm{0.370^*}$ &$\bm{0.429^*}$ &$\bm{0.522^*}$ &$\bm{0.689^*}$ &$\bm{0.428^*}$ &$\bm{0.466^*}$ &$\bm{0.557^*}$ &$\bm{0.695^*}$ &$\bm{0.344^*}$ &$\bm{0.386^*}$ &$\bm{0.476^*}$ &$\bm{0.581^*}$\\
         \midrule
         SASRec &0.348 &0.411 &0.508 &0.678 &0.412 &0.456 &0.543 &0.667 &0.322 &0.357 &0.439 &0.558\\
         CASR-SASRec &0.357 &0.415 &0.518 &0.685 &0.420 &0.461 & 0.549 &0.673 &0.335 &0.365 &0.450 &0.575\\
         CCR-SASRec &$\bm{0.376^*}$ &$\bm{0.427^*}$ &$\bm{0.531^*}$ &$\bm{0.701^*}$ &$\bm{0.438^*}$ &$\bm{0.479^*}$ &$\bm{0.562^*}$ &$\bm{0.699^*}$ &$\bm{0.358^*}$ &$\bm{0.390^*}$ &$\bm{0.471^*}$ &$\bm{0.592^*}$\\
         \midrule
         NCR &0.359 &0.412 &0.514 &0.680 &0.415 &0.457 &0.551 &0.673 &0.332 &0.366 &0.441 &0.557\\
         CASR-NCR &0.362 &0.419 &0.518 &0.689 &0.417 &0.458 &0.555 &0.682 &0.339 &0.374 &0.451 &0.569\\
         CCR-NCR &$\bm{0.376^*}$ &$\bm{0.434^*}$ &$\bm{0.535^*}$ &$\bm{0.705^*}$ &$\bm{0.433^*}$ &$\bm{0.472^*}$ &$\bm{0.568^*}$ &$\bm{0.702^*}$ &$\bm{0.354^*}$ &$\bm{0.395^*}$ &$\bm{0.469^*}$ &$\bm{0.588^*}$\\
         \bottomrule
    \end{tabular}
    \caption{Experimental results on Normalize Discounted Cumulative Gain (NDCG) and Hit Ratio (HR). For each model, we present the performance of the original model, the results of applying implicit counterfactual data augmentation method CASR on the model, and the results of applying our CCR method on the model. 
    Bold numbers represent best performance. We use * to indicate that the performance is significant better than other baselines. The significance is at 0.05 level on paired t-test.}
    \label{Table:prediction}
    \vspace{-20pt}
\end{table*}

\subsection{Implementation Details}
The learning rate is searched in [0.0001,0.001,0.01,0.1] for all methods. We apply ReLU as the activation function. For all methods, the embedding size is 64. We optimize the methods using mini-batch \cite{kingma2014adam} and the batch size is 128. The hyper-parameter $\alpha$ is searched in [$10^{-3}$,$10^{-2}$,$10^{-1}$,1,$10^{1}$,$10^{2}$,$10^{3}$], and finally set to 10 for the results we report in the paper. The confidence parameter $\kappa$ is searched from 0 to 1. The influence of different $\kappa$ on the performance will be discussed in the following experiments. We tune each model's parameters to its own best performance on the validation set. For both CASR (implicit counterfactual data augmentation) and CCR (explicit counterfactual data augmentation), we generate one counterfactual example for each sequence in the training set.

\subsection{Evaluation Metrics}
We use Normalized Discounted Cumulative Gain at rank $K$ (NDCG@K) and Hit Ratio at rank $K$ (HR@K) to evaluate recommendation performance. To evaluate the explanation performance, we use Probability of Necessity (PN), Probability of Sufficiency (PS) and their harmonic mean $\rm{F_{NS}} = \frac{2\cdot \rm{PN} \cdot \rm{PS}}{\rm{PN}+\rm{PS}}$. For each user-item pair in the validation set and the test set, we randomly sample 100 irrelevant items and rank all of these 101 items for recommendation.

\begin{table*}[t]
    \centering
    \setlength{\tabcolsep}{4pt}
    \begin{tabular}{l|r|r|r|r|r|r|r|r|r|r|r|r|r|r|r|r|r|r}
    \toprule
         Dataset & \multicolumn{6}{|c|}{ML100K} & \multicolumn{6}{c|}{Movies \& TV}& \multicolumn{6}{c}{Electronics}\\
         \midrule
         Top N  & \multicolumn{3}{|c|}{N=1} & \multicolumn{3}{|c|}{N=5} & \multicolumn{3}{|c|}{N=1} & \multicolumn{3}{|c|}{N=5} & \multicolumn{3}{|c|}{N=1} & \multicolumn{3}{|c}{N=5}\\
         \midrule
         Metric &\footnotesize PN\% &\footnotesize PS\% &\footnotesize $\rm{F_{NS}}$\% &\footnotesize PN\% &\footnotesize PS\% &\footnotesize $\rm{F_{NS}}$\% &\footnotesize PN\% &\footnotesize PS\% &\footnotesize $\rm{F_{NS}}$\% &\footnotesize PN\% &\footnotesize PS\% &\footnotesize $\rm{F_{NS}}$\% &\footnotesize PN\% &\footnotesize PS\% &\footnotesize $\rm{F_{NS}}$\% &\footnotesize PN\% &\footnotesize PS\% &\footnotesize $\rm{F_{NS}}$\%\\
         \midrule
        CASR-STAMP & 25.6 & 7.7 & 11.8 & 22.1 & 12.6 & 11.8
         & 29.2 & 17.8 & 22.2 & 18.4 & 17.1 & 17.7 
         & 30.9 & 8.7 & 13.6 & 24.9 &12.1 & 16.3\\
         CASR-GRU4Rec & 21.7 & 8.2 & 11.9 & 18.4 & 13.1 & 15.3
         & 23.9 & 3.5 & 6.1 & 19.8 & 8.4 & 11.8 
         & 22.4 & 9.9 & 13.8 & 20.6 &14.0 & 16.7\\
         CASR-SASRec & 23.9 & 9.3 & 13.4 & 19.3 & 13.5 & 15.8
         & 25.7 & 18.2 & 21.3 & 17.4 & 18.9 & 18.1 
         & 25.3 & 10.2 & 14.5 & 19.4 &14.5 & 16.6\\
         CASR-NCR & 17.2 & 32.5 & 22.5 & 14.7 & 36.8 & 21.0
         & 19.4 & 38.6 & 25.8 & 10.8 & 44.3 & 17.4
         & 19.9 & 39.0 & 26.3 & 16.3 &40.3 & 23.2\\
         \midrule
         CountER-STAMP & $\textbf{53.2}$ & 17.0 & 25.8 & $\textbf{38.3}$ & 26.4 & 31.2
         & $\textbf{58.6}$ & 36.8 & 45.2 & $\textbf{47.9}$ & 43.8 & 45.8
         & $\textbf{59.6}$ & 18.6 & 28.3 & $\textbf{48.1}$ & 27.5 & 34.9\\
         CountER-GRU4Rec & 40.8 & 19.1 & 26.0 & 34.5 & 29.5 & 31.8
         & 46.1 & 6.9 & 12.0 & 41.3 & 14.9 & 21.9 
         & 43.5 & 19.2 & 26.6 & 39.7 &29.6 & 33.9\\
         CountER-SASRec & 45.3 & 21.9 & 29.5 & 36.0 & 30.1 & 32.7
         & 48.3 & 39.7 & 43.5 & 40.5 & 45.5 & 42.8 
         & 50.7 & 23.9 & 32.4 & 41.2 &36.6 & 38.7\\
         CountER-NCR & 34.7 & 52.4 & 41.7 & 28.1 & 54.5 & 37.1
         & 42.7 & 53.7 & 47.5 & 34.9 & 59.0 & 43.8 
         & 36.7 & 52.3 & 43.2 & 32.2 &57.8 & 41.3\\
         \midrule
         CCR & 42.1 & $\textbf{60.3}$ & $\textbf{49.6}$ & 36.1 & $\textbf{66.7}$ & $\textbf{46.8}$
         & 50.1 & $\textbf{64.7}$ & $\textbf{56.5}$ & 41.9 & $\textbf{73.2}$ & $\textbf{53.3}$
         & 45.8 & $\textbf{74.8}$ & $\textbf{56.8}$ & 41.0 & $\textbf{79.6}$ & $\textbf{54.1}$\\
         \bottomrule
    \end{tabular}
    \caption{Results on PN, PS and $\rm{F_{NS}}$. Bold numbers are best performance. All numbers are percentage numbers with $\%$ omitted. When CCR achieves the best result, its improvements against the best baseline are significant at $p < 0.01$.}
    \label{Table:explain}
    \vspace{-20pt}
\end{table*}

\begin{figure*}[t]
\centering
\subfigure{
\begin{minipage}[t]{0.3\linewidth}
\centering
\begin{tikzpicture}[scale=0.5] 
    \begin{axis}[
        ylabel near ticks,
        xlabel near ticks,
        xlabel= CCR-NCR results on $ML100K$,
        ylabel=$HR@10$,
        ymin=0.6, ymax=0.75,
        xtick = {1,2,3,4,5,6,7,8,9,10,11},
        xticklabels = {0,0.1,0.2,0.3,0.4,0.5,0.6,0.7,0.8,0.9,1.0}
        ]
    \addplot[smooth,mark=*,blue,dash pattern=on 1pt off 3pt on 3pt off 3pt] plot coordinates {
        (1,0.69)
        (2,0.69)
        (3,0.69)
        (4,0.69) 
        (5,0.6905)
        (6,0.6998)
        (7,0.7091)
        (8,0.7067)
        (9,0.7002)
        (10,0.6952)
        (11,0.6905)

    };
    \end{axis}
    
    \begin{axis}[
        axis y line*=right,
        axis x line*=none,
        ylabel near ticks,
        ylabel=$NDCG@10$,
        ymin=0.4, ymax=0.47,
        xtick = {1,2,3,4,5,6,7,8,9,10,11},
        xticklabels = {0,0.1,0.2,0.3,0.4,0.5,0.6,0.7,0.8,0.9,1.0},
        axis x line*=none
        ]
    \addplot[smooth,mark=*,red,dash pattern=on 1pt off 3pt on 3pt off 3pt] plot coordinates {
        (1,0.4198)
        (2,0.4198)
        (3,0.4198)
        (4,0.4198) 
        (5,0.4201)
        (6,0.4291)
        (7,0.4348)
        (8,0.4309)
        (9,0.4302)
        (10,0.4296)
        (11,0.4293)
    };
    \end{axis}

    \end{tikzpicture}
\end{minipage}%
}%
\hspace{-10pt}%
\subfigure{
\begin{minipage}[t]{0.3\linewidth}
\centering
\begin{tikzpicture}[scale=0.5] 
    \begin{axis}[
        ylabel near ticks,
        xlabel near ticks,
        xlabel= CCR-NCR results on $Movies \& TV$,
        ylabel=$HR@10$,
        ymin=0.65, ymax=0.73,
        xtick = {1,2,3,4,5,6,7,8,9,10,11},
        xticklabels = {0,0.1,0.2,0.3,0.4,0.5,0.6,0.7,0.8,0.9,1.0}
        ]
    \addplot[smooth,mark=*,blue,dash pattern=on 1pt off 3pt on 3pt off 3pt] plot coordinates {
        (1,0.6912)
        (2,0.6912)
        (3,0.6912)
        (4,0.6912) 
        (5,0.6912)
        (6,0.6954)
        (7,0.7005)
        (8,0.70)
        (9,0.7033)
        (10,0.6951)
        (11,0.6902)

    };
    \end{axis}
    
    \begin{axis}[
        axis y line*=right,
        axis x line*=none,
        ylabel near ticks,
        ylabel=$NDCG@10$,
        ymin=0.46, ymax=0.50,
        xtick = {1,2,3,4,5,6,7,8,9,10,11},
        xticklabels = {0,0.1,0.2,0.3,0.4,0.5,0.6,0.7,0.8,0.9,1.0},
        axis x line*=none
        ]
    \addplot[smooth,mark=*,red,dash pattern=on 1pt off 3pt on 3pt off 3pt] plot coordinates {
        (1,0.4701)
        (2,0.4701)
        (3,0.4701)
        (4,0.4701) 
        (5,0.4701)
        (6,0.4721)
        (7,0.473)
        (8,0.4726)
        (9,0.4742)
        (10,0.469)
        (11,0.4667)
    };
    \end{axis}
    \end{tikzpicture}%
    \end{minipage}%
}%
\hspace{-15pt}%
\subfigure{\begin{minipage}[t]{0.3\linewidth}
\centering

\begin{tikzpicture}[scale=0.5] 
    \begin{axis}[
        ylabel near ticks,
        xlabel near ticks,
        xlabel= CCR-NCR results on $Electronics$,
        ylabel=$HR@10$,
        ymin=0.54, ymax=0.60,
        xtick = {1,2,3,4,5,6,7,8,9,10,11},
        xticklabels = {0,0.1,0.2,0.3,0.4,0.5,0.6,0.7,0.8,0.9,1.0}
        ]
    \addplot[smooth,mark=*,blue,dash pattern=on 1pt off 3pt on 3pt off 3pt] plot coordinates {
        (1,0.5693)
        (2,0.5693)
        (3,0.5693)
        (4,0.5693) 
        (5,0.5707)
        (6,0.5761)
        (7,0.5789)
        (8,0.5848)
        (9,0.581)
        (10,0.5753)
        (11,0.5678)

    };
    \end{axis}
    
    \begin{axis}[
        axis y line*=right,
        axis x line*=none,
        ylabel near ticks,
        ylabel=$NDCG@10$,
        ymin=0.37, ymax=0.43,
        xtick = {1,2,3,4,5,6,7,8,9,10,11},
        xticklabels = {0,0.1,0.2,0.3,0.4,0.5,0.6,0.7,0.8,0.9,1.0},
        axis x line*=none
        ]
    \addplot[smooth,mark=*,red,dash pattern=on 1pt off 3pt on 3pt off 3pt] plot coordinates {
        (1,0.3802)
        (2,0.3802)
        (3,0.3802)
        (4,0.3802) 
        (5,0.3856)
        (6,0.3882)
        (7,0.3902)
        (8,0.3984)
        (9,0.3956)
        (10,0.3842)
        (11,0.376)
    };
    \end{axis}
     
    \end{tikzpicture}%
\end{minipage}%
}
\vspace{-15pt}
\caption{Performance on HR@10 (Blue Line) and NDCG@10 (Red Line) on different $\kappa$ with different datasets.}
\label{figure: kappa}
\vspace{-15pt}
\end{figure*}
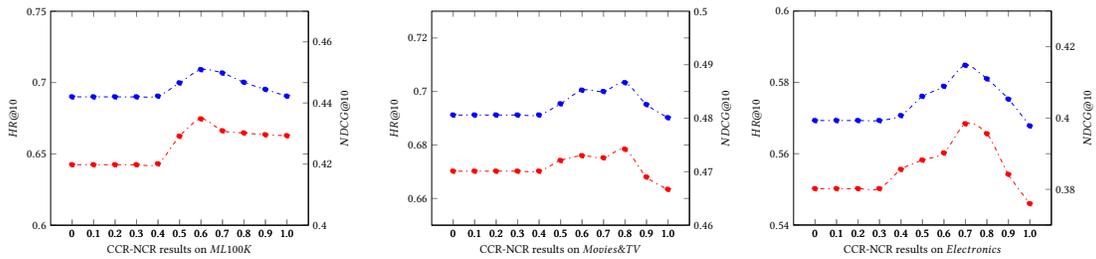

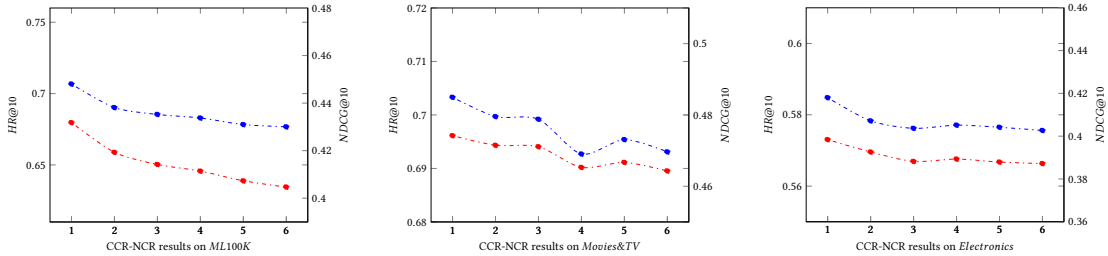
\begin{figure*}[t]
\centering
\subfigure{
\begin{minipage}[t]{0.3\linewidth}
\centering
\begin{tikzpicture}[scale=0.5] 
    \begin{axis}[
        ylabel near ticks,
        xlabel near ticks,
        xlabel= CCR-NCR results on $ML100K$,
        ylabel=$HR@10$,
        ymin=0.61, ymax=0.76,
        xtick = {1,2,3,4,5,6},
        xticklabels = {1,2,3,4,5,6}
        ]
    \addplot[smooth,mark=*,blue,dash pattern=on 1pt off 3pt on 3pt off 3pt] plot coordinates {
        (1,0.7067)
        (2,0.6902)
        (3,0.6854)
        (4,0.6829) 
        (5,0.6783)
        (6,0.6767)

    };
    \end{axis}
    
    \begin{axis}[
        axis y line*=right,
        axis x line*=none,
        ylabel near ticks,
        ylabel=$NDCG@10$,
        ymin=0.39, ymax=0.48,
        xtick = {1,2,3,4,5,6},
        xticklabels = {1,2,3,4,5,6},
        axis x line*=none
        ]
    \addplot[smooth,mark=*,red,dash pattern=on 1pt off 3pt on 3pt off 3pt] plot coordinates {
        (1,0.4318)
        (2,0.4193)
        (3,0.4142)
        (4,0.4114) 
        (5,0.4073)
        (6,0.4047)

    };
    \end{axis}

    \end{tikzpicture}
\end{minipage}%
}%
\hspace{-10pt}%
\subfigure{
\begin{minipage}[t]{0.3\linewidth}
\centering
\begin{tikzpicture}[scale=0.5] 
    \begin{axis}[
        ylabel near ticks,
        xlabel near ticks,
        xlabel= CCR-NCR results on $Movies \& TV$,
        ylabel=$HR@10$,
        ymin=0.68, ymax=0.72,
        xtick = {1,2,3,4,5,6},
        xticklabels = {1,2,3,4,5,6}
        ]
    \addplot[smooth,mark=*,blue,dash pattern=on 1pt off 3pt on 3pt off 3pt] plot coordinates {
        (1,0.7033)
        (2,0.6997)
        (3,0.6992)
        (4,0.6927) 
        (5,0.6954)
        (6,0.6931)

    };
    \end{axis}
    
    \begin{axis}[
        axis y line*=right,
        axis x line*=none,
        ylabel near ticks,
        ylabel=$NDCG@10$,
        ymin=0.45, ymax=0.51,
        xtick = {1,2,3,4,5,6},
        xticklabels = {1,2,3,4,5,6},
        axis x line*=none
        ]
    \addplot[smooth,mark=*,red,dash pattern=on 1pt off 3pt on 3pt off 3pt] plot coordinates {
        (1,0.4742)
        (2,0.4715)
        (3,0.4711)
        (4,0.4653) 
        (5,0.4667)
        (6,0.4643)

    };
    \end{axis}
     
    \end{tikzpicture}%

    \end{minipage}%
}%
\hspace{-10pt}%
\subfigure{\begin{minipage}[t]{0.3\linewidth}
\centering

\begin{tikzpicture}[scale=0.5] 
    \begin{axis}[
        ylabel near ticks,
        xlabel near ticks,
        xlabel= CCR-NCR results on $Electronics$,
        ylabel=$HR@10$,
        ymin=0.55, ymax=0.61,
        xtick = {1,2,3,4,5,6},
        xticklabels = {1,2,3,4,5,6}
        ]
    \addplot[smooth,mark=*,blue,dash pattern=on 1pt off 3pt on 3pt off 3pt] plot coordinates {
        (1,0.5848)
        (2,0.5783)
        (3,0.5762)
        (4,0.5771) 
        (5,0.5765)
        (6,0.5756)

    };
    \end{axis}
    
    \begin{axis}[
        axis y line*=right,
        axis x line*=none,
        ylabel near ticks,
        ylabel=$NDCG@10$,
        ymin=0.36, ymax=0.46,
        xtick = {1,2,3,4,5,6},
        xticklabels = {1,2,3,4,5,6},
        axis x line*=none
        ]
    \addplot[smooth,mark=*,red,dash pattern=on 1pt off 3pt on 3pt off 3pt] plot coordinates {
        (1,0.3984)
        (2,0.3926)
        (3,0.3882)
        (4,0.3893) 
        (5,0.3879)
        (6,0.3872)

    };
    \end{axis}
     
    \end{tikzpicture}%
\end{minipage}%
}%
\vspace{-15pt}
\caption{Performance on HR@10 (Blue Line) and NDCG@10 (Red Line) with different iterations.}
\label{figure: iteration}
\vspace{-10pt}
\end{figure*}

\subsection{Compatible with the Baselines}
One issue in comparison with baselines is that our framework can generate explanations in the counterfactual generation progress while the baseline methods can not. We use two approaches to make the baselines compatible for explanation evaluation.

First, we apply CASR on each of the four recommendation models to generate explanations. Since CASR manually selects one item to intervene, we intervene each item in each example's history and select the one that achieves the highest $\rm{F_{NS}}$ score as the CASR explanation.
Second, to get stronger explanations for each model, we use our framework as a guideline to tell the baseline methods how many items they should use to generate explanations. Then, we apply the counterfactual explainable recommendation (CountER) framework \cite{tan2021counterfactual} to each of the recommendation model (STAMP, GRU4Rec, SASRec, NCR) to generate explanations for them: based on the number of items, we search all of the combinations of users' history items as candidate explanations and take the one that gives highest $\rm{F_{NS}}$ score as the explanation of the model under CountER framework. 
Finally, since the CCR framework works in the ``augment once, apply to all'' paradigm and directly produces explicit explanation with NCR sampler during data augmentation, we directly take its output explanation for evaluation.

\subsection{\mbox{Performance on Recommendation}}
The experimental results on NDCG and Hit Ratio (HR) are shown in Table \ref{Table:prediction}. 
Based on the results, we have following observations.

First and most importantly, compared with the original model and the model under implicit counterfactual data augmentation, our CCR framework achieves significantly better performance than the baseline methods on all of these three datasets. Compared with the original model, CCR can get better results based on the the generated explicit counterfactual data, which alleviates the data scarcity and encodes informative examples into the training dataset. Compared with the implicit counterfactual data augmentation method CASR, the explicit counterfactual data generated by CCR are more effective since CCR takes advantages of the explicit feedback. 


\subsection{Performance on Explanation}
The experimental results on explanation are shown in Table \ref{Table:explain}. 
Based on the experiment results, we have following observations.

In terms of the overall explanation performance ($\rm{F_{NS}}$), CountER-based explanations are better than CASR-based explanations. This is understandable since CountER uses the optimal number of items from CCR to generate explanations while CASR selects one and only one item as explanation. Furthermore, by considering explicit feedback, CCR generates even better explanations than CountER-based methods. 

Besides, by considering explicit feedback, the CCR explanations are better than CASR explanations on both PN and PS and thus better overall explanation quality $\rm{F_{NS}}$. 
This means that the CCR augmented examples have higher quality since they are more sufficient and necessary, and when these better examples are used to augment the dataset, it helps CCR to achieve better recommendation performance than CASR.

\subsection{Impact of Hyper-Parameters}

\subsubsection{\textbf{Impact of $\kappa$}}
In the generation process of the explicit counterfactual data, we have a confidence parameter $\kappa$. We accept the generated counterfactual data only when the ranking score of the data is larger than $\kappa$. The results on the influence of $\kappa$ is shown in Figure \ref{figure: kappa}. From the figure, we can see that our framework will have the best performance when $\kappa$ is set around 0.7. When $\kappa$ is very small, it does not change the performance of the framework because all of the generated data can pass the constraint of $\kappa$. When $\kappa$ is too big, the performance will decrease because only a few counterfactual data can pass through the constraint and thus we cannot get enough counterfactual data to re-optimize the anchor model.

\vspace{-1ex}
\subsection{Impact of Iterative Re-optimization}
Since the re-optimized anchor model $\bm{\mathcal{A}^\prime}$ achieves better performance than the original anchor model $\bm{\mathcal{A}}$, a natural idea is that if we use the re-optimized anchor model to generate a set of new augmented data and optimize $\bm{\mathcal{A}^\prime}$ again to $\bm{\mathcal{A}^{\prime\prime}}$ in a boosting way, whether the performance can be even better. We use NCR as an example anchor model to test the idea.
Unfortunately, as we can see in Figure \ref{figure: iteration}, the performance on HR and NDCG decreases with the number of rounds of iterative re-optimization. 
This observation is consistent with previous research \cite{wang2021counterfactual}. The reason is that since the sampler is not perfectly accurate, the generated counterfactual examples can contain noise and such noise is learned into the anchor model. As a result, multiple rounds of augmentation and re-optimization may propagate such noise and thus decrease the performance. This means that even though data augmentation can improve the model performance, it cannot boost the model performance infinitely.

\vspace{-1ex}
\section{Conclusion}
\label{sec:conclusions}

In this paper, we propose a Counterfactual Collaborative Reasoning (CCR) framework, which integrates the power of logical reasoning and counterfactual reasoning and generates explicit counterfactual data to enhance the performance of sequential recommendation models. Experiments on three real-world datasets verified the effectiveness of the framework. Furthermore, a unique advantage of the CCR framework is that it can also generate explicit counterfactual explanations to better understand the user behavior sequence. In this work, we take recommender system as an example to explore the joint ability of logical and counterfactual reasoning, which are two important types of reasoning abilities for machine learning. On the other hand, they can also be considered to improve other intelligent tasks beyond recommendation, such as vision and language processing tasks, which we will explore in the future.
\\

\textbf{Acknowledgment}. This work was supported in part by NSF IIS 1910154, 2007907, 2046457, 2127918 and CCF 2124155. Any opinions, findings, conclusions
or recommendations expressed in this material are those of the
authors and do not necessarily reflect those of the sponsors.


\bibliographystyle{ACM-Reference-Format}
\balance
\bibliography{reference.bib}

\end{document}